\documentclass[conference]{IEEEtran}
\IEEEoverridecommandlockouts
\usepackage{cite}
\usepackage{amsmath,amssymb,amsfonts}
\usepackage{algorithmic}
\usepackage{graphicx}
\usepackage{textcomp}
\usepackage{xcolor}
\def\BibTeX{{\rm B\kern-.05em{\sc i\kern-.025em b}\kern-.08em
    T\kern-.1667em\lower.7ex\hbox{E}\kern-.125emX}}

\usepackage{bbding}
\usepackage{amsmath}
\usepackage{svg}
\usepackage{multirow}
\usepackage{gensymb}
\usepackage{hyperref}

\begin{document}

\title{Embedding Compression Distortion \\in Video Coding for Machines}

\author{
    \IEEEauthorblockN{
        Yuxiao Sun$^{1}$, 
        Yao Zhao$^{1}$, 
        Meiqin Liu$^{\ast, 1}$, 
        Chao Yao$^{2}$, 
        and Weisi Lin$^{3}$
        \thanks{
            $^{\ast}$Corresponding author
        }
        \thanks{
            This work is supported by the National Natural Science Foundation of China (62120106009, 62372036, U24B20179, U22A2022, 62332017).
        }
    }
    \IEEEauthorblockA{
    \text{$^{1}$Beijing Jiaotong University, China} \\
    \text{$^{2}$University of Scienece and Technology Beijing, China} \\
    \text{$^{3}$Nanyang Technological University, Singapore}\\
    \text{\{yuxiaosun, yzhao, mqliu\}@bjtu.edu.cn, yaochao@ustb.edu.cn, wslin@ntu.edu.sg} \\
    }
}


\maketitle

\begin{abstract}
Currently, video transmission serves not only the Human Visual System (HVS) for viewing but also machine perception for analysis. However, existing codecs are primarily optimized for pixel-domain and HVS-perception metrics rather than the needs of machine vision tasks. To address this issue, we propose a Compression Distortion Representation Embedding (CDRE) framework, which extracts machine-perception-related distortion representation and embeds it into downstream models, addressing the information lost during compression and improving task performance. Specifically, to better analyze the machine-perception-related distortion, we design a compression-sensitive extractor that identifies compression degradation in the feature domain. For efficient transmission, a lightweight distortion codec is introduced to compress the distortion information into a compact representation. Subsequently, the representation is progressively embedded into the downstream model, enabling it to be better informed about compression degradation and enhancing performance. Experiments across various codecs and downstream tasks demonstrate that our framework can effectively boost the rate-task performance of existing codecs with minimal overhead in terms of bitrate, execution time, and number of parameters. Our codes and supplementary materials are released in \url{https://github.com/Ws-Syx/CDRE/}. 

\end{abstract}


\section{Introduction}
\label{sec:intro}

Digital video plays a significant role in our daily lives, accounting for a large proportion of data traffic. Increasingly, video transmission serves not only human visual system (HVS) but also downstream machine vision tasks, such as surveillance video analysis and facial recognition. However, most existing codecs focus on pixel-domain and HVS-related metrics~\cite{dcvc_dc, dcvc_sdd, li2021dcvc, zhang2024tlvc, cheng2020} and introduce distortions that harm the accuracy of these downstream tasks, thereby damaging overall rate-task performance. This issue is regarded as a challenge within the field of Video Coding for Machines (VCM).

To address the performance degradation in machine vision tasks caused by compression distortion, a straightforward approach is restoring frames for downstream machine perception with a restoration model\cite{ddnet, dong2015cir, jiang2024cir}. 
Another simple yet effective solution is to optimize the machine vision model on compressed datasets, which makes the model better handle distorted inputs. However, these methods only generalize common distortion patterns and struggle to precisely capture the specific distortion encountered by each input frame. 
Moreover, introducing additional bitstreams has proven effective, such as maintaining semantic consistency between compressed and original images~\cite{tian2023smc, tian2024smc++} and enhancing edge details~\cite{tian2022benchmark}. However, a significant burden is introduced due to the increase in parameters and execution time.

\begin{figure}[t]
    \centering
    \includegraphics[width=0.99\columnwidth]{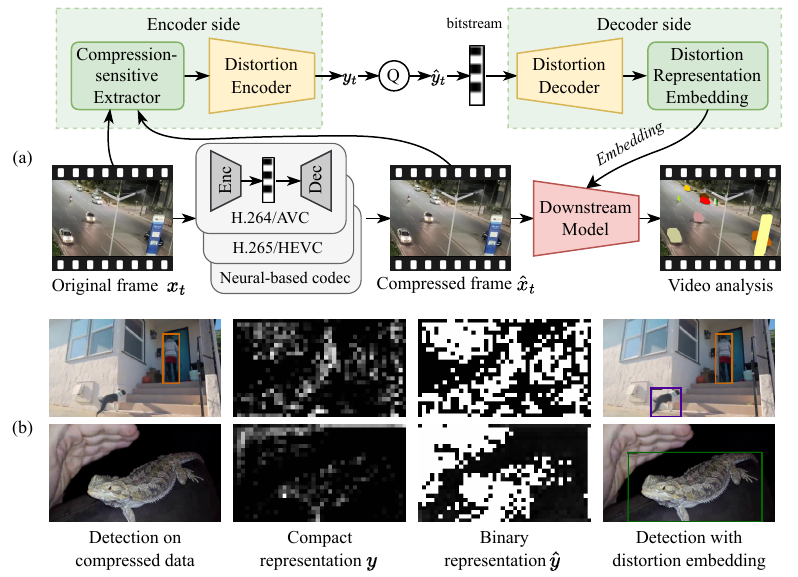}
    \vspace{-20pt}
    \caption{ (a) Overview of our proposed CDRE framework, which extracts, compresses, transmits, and embeds machine-perception-related distortion representations. It helps the downstream model be aware of specific distortions in the input data, improving task accuracy.  (b) Visualization of compact and binary representation of machine-perception-related distortion. Detection accuracy increases after distortion representation embedding. }
    \label{figure_intro_merge}
     \vspace{-20pt}
\end{figure}

\begin{figure*}[!h]
    \centering
    \includegraphics[width=0.98\textwidth]{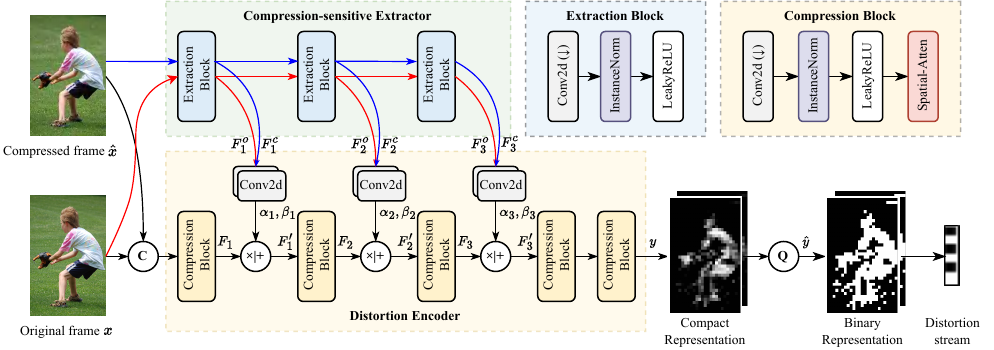}
    \vspace{-12pt}
    \caption{The process of distortion representation extraction and compression at the encoder side. $C$ denotes channel-wise concatenation, $Q$ denotes binary quantization, and $\times|+$ denotes linear modulation operations. Both the compressed image and the original image are input into the compression-sensitive extractor, yielding multi-level features $F^o_{1,2,3}$ and $F^c_{1,2,3}$ to identify distortions in the feature domain. Subsequently, under the modulation of the aforementioned features, the distortion encoder extracts and compresses the distortions into a compact representation, which is then quantized. }
    \label{figure_encoder}
    \vspace{-16pt}
\end{figure*}

Inspired by prompt learning~\cite{vpt}, we reinterpret the prompt as the compression distortion information of the current input, represented through an auxiliary bitstream. The lightweight Compression Distortion Representation Embedding (CDRE) framework is proposed to further improve the rate-task performance of existing codecs without introducing significant computational and bitrate overhead. This framework detects and compresses compression-induced distortion and fuses it into the machine perception process.
Due to the down-sampling and extraction in the backbone of machine perception models, pixel-domain texture and HVS-related details are often diminished, making pixel-domain and HVS-related quality less relevant to downstream performance. Instead of focusing on improving the quality of compressed frames, our approach embeds distortion information directly into the downstream inference process. By effectively incorporating distortion information in the feature domain, downstream models achieves better performance, as illustrated in Fig.~\ref{figure_intro_merge}.


In particular, a compression-sensitive extractor is introduced to identify compression degradation and mining machine-perception-related distortions in the feature domain. To effectively transmit the distortion information to the decoder side without putting extra burden on the encoder side, we employ a lightweight distortion codec to compress and quantize the distortion into a binary representation. After transmission, the representation is gradually decoded and embedded at multiple scales into the backbone of the downstream model by a Distortion Representation Embedding Module, enabling it to perform inference with prior knowledge of the current compression distortion. The effectiveness of the CDRE framework is validated on various existing codecs, including both standard and neural-based codecs. Experimental results indicate that our CDRE effectively enhances rate-task performance in object detection, semantic segmentation, and video instance segmentation. Additionally, only minimal overhead is introduced by CDRE in terms of bitrate, execution time, and number of parameters, making it suitable for scenarios with limited encoder-side resources.

Our contributions are summarized as follows: 

\begin{itemize}
    \item We propose a lightweight Compression Distortion Representation Embedding (CDRE) framework, which enhances rate-task performance in VCM scenarios by extracting and embedding machine-perception-related distortion representation in the feature domain. 
    
    \item We introduce a compression-sensitive extractor for analyzing the feature-domain distortion and a lightweight distortion codec to compress the distortion representation with minimal computation and bitrate overhead.  
    
    \item We design a multi-scale distortion representation embedding module to progressively integrate distortion information into downstream models, effectively utilizing this information as a prior during the inference process.
    
\end{itemize}


\section{Method}

\subsection{Framework Overview}

The pipeline of the CDRE framework is shown in Fig.~\ref{figure_intro_merge}~(b). It contains three main processes: video compression by existing codecs, distortion representation extraction and compression, and distortion representation embedding. Firstly, videos are compressed, transmitted, and reconstructed through existing codecs. Once video transmission is completed, the compression distortion is extracted by a compression-sensitive extractor and compressed into a compact representation $y$ using a distortion encoder. Then it is quantized into binary representation $\hat{y}$ and transmitted to the decoder side with a minimal bitstream and computation overhead. On the decoder side, the representation $\hat{y}$ is reconstructed and embedded into the backbone of downstream tasks in a multi-scale manner, enabling downstream models to better counteract compression distortions.

\subsection{Distortion Representation Extraction}

Existing standard and neural-based codecs typically reduce pixel-domain and HVS-perception-related distortion using metrics such as PSNR, MS-SSIM, and LPIPS~\cite{hevc_overview, dcvc_dc, li2021dcvc, cheng2020, zhang2024tlvc, dcvc_sdd}. However, the pixel-domain metric PSNR and the HVS-related metrics MS-SSIM and LPIPS do not exhibit a strong correlation with machine perception, leading to suboptimal rate-task performance, as detailed in the supplementary materials. This issue stems from that the down-sampling and feature extraction of downstream backbones discard pixel-domain and HVS-related details, reducing the impact of pixel-domain and HVS-related quality on task accuracy.
While distortions related to the HVS, such as motion blur and blocking artifacts, are well-defined, machine-perception distortions lack a clear definition.
To overcome this limitation, we propose investigating machine-perception-related distortions by shifting the focus from pixel-domain metrics to feature-domain analysis. Specifically, we defined feature-domain distortion as:

\begin{equation}
\begin{aligned}
    D = - cosine(F^o, F^c)
\end{aligned}
\label{eq_distortion}
\end{equation}

\noindent where $D$ denotes machine-perception-related distortion in the feature domain, and $cosine(\cdot)$ denotes cosine similarity. $F^o$ and $F^c$ are original features and compression-distorted features mentioned later, respectively.

\begin{figure}[t]
    \centering
    \includegraphics[width=0.95\linewidth]{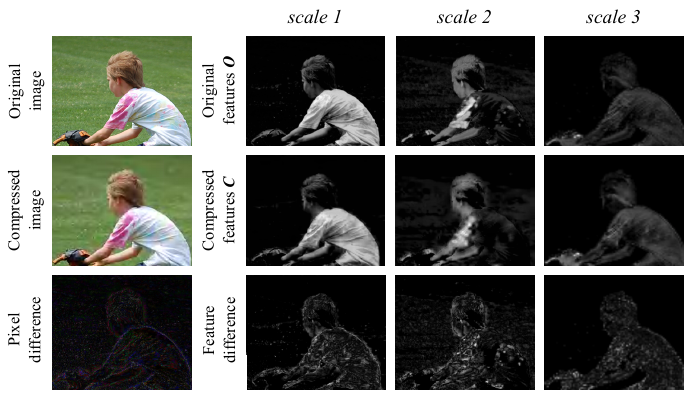}
    \vspace{-10pt}
    \caption{Visual example from MS-COCO-2017 dataset. White means a higher value in the feature map. Since the feature distance between original and compressed images is amplified at three distinct scales, distortion is more easily discernible in the feature domain than in the pixel domain.}
    \label{figure_sensitive_feature}
    \vspace{-16pt}
\end{figure}

To effectively perceive distortions in the feature domain, we introduce a lightweight compression-sensitive extractor. 
As shown in Fig.~\ref{figure_encoder}, the extractor contains three simple extraction blocks, avoiding computational burdens on the encoder side. As dedicated in Fig.~\ref{figure_sensitive_feature}, it amplifies the compression degradation in the feature domain between the original and compressed frames, thereby enhancing the focus on areas where feature distortion occurs during the subsequent extraction and compression process. Specifically, the original image $x$ and compressed image $\hat{x}$ are fed into the extractor separately to obtain features $F^o$ and $F^c$, enabling a hierarchical perception of distortion, as detailed by the following equations.

\begin{equation}
\begin{aligned}
    F^o_i &= \mathrm{LeakyReLU}(\mathrm{Norm}(\mathrm{Conv}(F^o_{i-1})) \\
    F^c_i &= \mathrm{LeakyReLU}(\mathrm{Norm}(\mathrm{Conv}(F^c_{i-1})) \\
\end{aligned}
\label{eq_sensitive_extractor}
\end{equation}

\noindent where $i$ is the layer-index, $\mathrm{Conv}(\cdot)$ denotes a convolutional layer, $\mathrm{Norm}(\cdot)$ denotes an Instance Normalization layer, and $\mathrm{LeakyReLU}(\cdot)$ denotes a LeakyReLU activation layer.

\subsection{Distortion Representation Compression}

To reduce the bitrate of transmitting distortion representation to downstream processes, we employ a VAE-like network~\cite{balle2016end} for distortion representation compression. It takes the original image $x$ and compressed image $\hat{x}$ as input and considers features $F^o$ and $F^c$ as compression condition. The compression process is dedicated to the following equation. 

\begin{equation}
F_d = \mathcal{D}(\mathcal{Q}(\mathcal{E}(x, \hat{x} | F^o, F^c))) 
\label{eq_codec}
\end{equation}

\noindent where $F_d$ is reconstructed distortion feature, $Q(\cdot)$ denotes binary quantization mentioned later, $\mathcal{E}(\cdot)$ and $\mathcal{D}(\cdot)$ denote distortion encoder and decoder, respectively. 

The distortion encoder down-samples the distortion information and squeezes it to a compact representation $y$ with lower channel and spatial dimensions. To better focus on the regions that are severely distorted and beneficial for downstream machine perception, features $F^o$ and $F^c$ are used for linear modulation during distortion representation encoding, as shown in equation~\ref{eq_modulation}. 

\vspace{-5pt}

\begin{equation}
\begin{aligned}
\alpha_i = \mathrm{Conv}(&F^o_i, F^c_i), \beta_i = \mathrm{Conv}(F^o_i, F^c_i) \\
F_i' &= \alpha_i \times F_i + \beta_i
\end{aligned}
\label{eq_modulation}
\end{equation}

\vspace{-5pt}

\noindent where $\alpha_i$ and $\beta_i$ are $i$-th scaling factor and shifting factor. $F_i$ is the output feature of $i$-th compression block in distortion encoder and $F_i'$ is distortion-modulated feature. 

Similar to the distortion encoder, the decoder contains convolutional layers for up-sampling and restoring the distortion representation. Notably, for CNN-based downstream backbones, the decoder reconstructs the binary representation $\hat{y}$ to the original spatial dimensions. In contrast, for Transformer-based backbones, the decoder reconstructs $\hat{y}$ to match the channel dimension of tokens in the downstream backbone. Details of the decoder are in supplementary materials. 

For lower bitrate consumption, the compact representation $y$ is quantized. Existing neural-based codecs typically quantize the floating-point matrix into integers and then use entropy models for entropy coding~\cite{dcvc_dc, li2021dcvc, cheng2020, lu2019dvc}. However, entropy models introduce heavy computation overhead and require extensive iterations for training. Instead, we introduce a simpler method. The value range of the compact representation is re-scaled to between 0 and 1 by the Sigmoid function. Then the existing quantization is simplified to binary quantization, as shown in Equation~\ref{eq5}. 

\begin{equation}
    \hat{y} = \lfloor Sigmoid(y) \rceil
    \label{eq5}
\end{equation}

\noindent where $y$ and $\hat{y}$ are compact representation and binary representation of distortion, respectively. $\lfloor \cdot \rceil$ denotes round opeartion.

\subsection{Distortion Representation Embedding}

\begin{figure}[t]
    \centering
    \includegraphics[width=0.98\columnwidth]{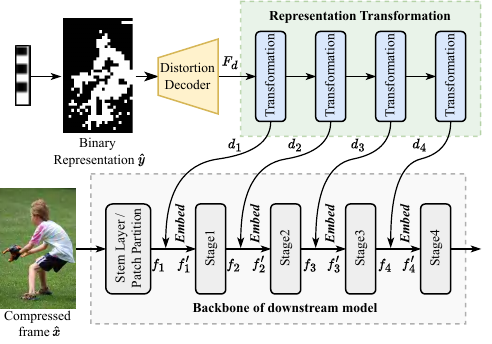}
    \vspace{-10pt}
    \caption{The process of distortion representation reconstruction and embedding at the decoder side. The representation is reconstructed by the distortion decoder. Then it is transferred and embedded into the backbone of the downstream model for better performance. The structure of the distortion decoder and distortion transformation module are detailed in supplementary materials. }
    \label{figure_decoder}
    \vspace{-16pt}
\end{figure}

Inspired by prompt learning~\cite{vpt}, we embed distortion features into the backbone of the downstream task model. This integration allows the model to remain aware of distortion information in the current input, enhancing its performance on downstream tasks. To fully utilize the distortion information, the distortion feature $F_d$ is further transformed into $d_{1,2,3,4}$ and embedded at four distinct scales, as shown in Fig.~\ref{figure_decoder}. Notably, the module ensures that the dimensions of the output feature $d_{1,2,3,4}$ align with each stage in the downstream backbone.

The distortion representation embedding operation integrates distortion information with the inference process. Given that CNN and Transformers are the most widely used backbones in machine vision, especially ResNet~\cite{faster_rcnn} and Swin-transformer~\cite{mask2former}, we propose two distinct embedding methods. For the CNN-based backbone, the process of distortion feature transformation and embedding is described by the following equation.

\begin{equation}
\begin{aligned}
    d_i &= \mathrm{ReLU}(\mathrm{Norm}(\mathrm{Conv}(d_{i-1}))) \\
    f_i' &= f_i + \mathrm{CA}(\mathrm{SA}(f_i, d_i)) 
\end{aligned}
\label{eq_cnn_embedding}
\end{equation}

\noindent where $\mathrm{CA}(\cdot)$ is channel-attention operation, $\mathrm{SA}(\cdot)$ is spatial-attention operation, $f_i$ represents $i$-th intermediate result in backbone, and $f_i'$ represents embedded feature. 

For the Transformer-based backbone, MLP is used for distortion feature transformation and cross-attention is used for distortion representation embedding, as detailed in the following equation. 

\begin{equation}
\begin{aligned}
    d_i &= \mathrm{MLP}(d_{i-1}) \\
    Q_i &= f_i W^Q, K_i = d_i W^K, V_i = d_i W^V \\
    f_i' &= f_i + Softmax(Q_i K_i^T / \sqrt{d}) V_i
\end{aligned}
\label{eq_vit_embedding}
\end{equation}

\noindent where $\mathrm{MLP}(\cdot)$ represents MLP with one hidden layer, $W^Q$, $W^K$ and $W^V$ represent the linear projections, and $d$ is the dimension of projected matrices. 

More details about distortion representation embedding is described in supplementary materials. 

\begin{figure*}[ht]
    \centering
    \includegraphics[width=0.95\textwidth]{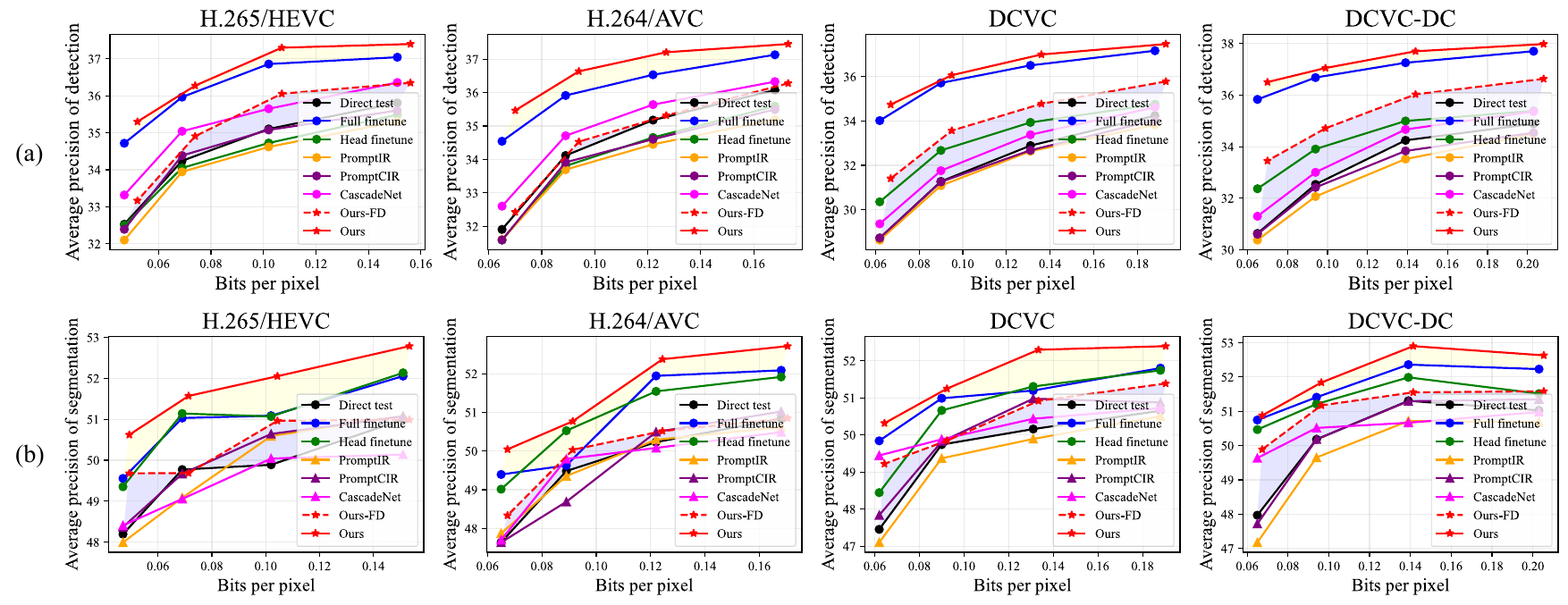}
    \vspace{-12pt}
    \caption{The rate-task performance of (a) object detection and (b) Video instance segmentation across various existing HVS-oriented codecs. ``Ours-FD'' represents CDRE modules are optimized but the downstream model is frozen, ``Ours'' represents CDRE modules and the downstream model are jointly optimized. The average precision on uncompressed data is 37.3\% (object detection) and 51.5\% (instance segmentation). }
    \label{curve_hvs}
     \vspace{-18pt}
\end{figure*}

\subsection{Loss Function}

Since the entire framework is designed for downstream machine perception, its primary optimization target is the performance of downstream tasks. Additionally, for better mining machine-perception-related distortion information, feature distance is amplified by cosine similarity. In summary, the loss function $
\mathcal{L}$ of the framework is as follows. 

\begin{equation}
    \mathcal{L} = \mathcal{L}_{task} + \lambda \sum_{i=1}^3 (1 + cosine(F^o_i, F^c_i))
    \label{eq7}
\end{equation}

\noindent where $\mathcal{L}_{task}$ is downstream task loss and $\lambda$ is balance weight.

\section{Experimental Results}

\subsection{Implementation Settings}

\subsubsection{Downstream Machine Vision Tasks} 

The proposed VCED framework is validated across three tasks with different granularities: video instance segmentation (Mask2Former\cite{mask2former}), person keypoints detection (Faster R-CNN\cite{keypoint_rcnn}), and object detection (Keypoint R-CNN\cite{faster_rcnn}), using YoutubeVIS-2019 (YTVIS2019)~\cite{mask2former} and MS-COCO-2017~\cite{faster_rcnn} datasets.

\subsubsection{Compression Methods}

Seven compression methods are used. For video compression, we employ standard codecs H.265/HEVC~\cite{hevc_overview}, H.264/AVC~\cite{avc_overview}, and neural-based high-performance video codecs DCVC~\cite{li2021dcvc} and DCVC-DC~\cite{dcvc_dc}. For image compression, we use standard codecs H.265/HEVC-Intra, H.264/AVC-Intra, JPEG-2000~\cite{jpeg_overview}, and neural-based codec Cheng-2020~\cite{cheng2020}. Furthermore, the VCM-related video codec SMC++~\cite{tian2024smc++} is included in our experiment. 

\subsubsection{Compared Methods}

We used two types of comparison methods. One method involves directly fine-tuning the downstream task model on compressed datasets, enabling models to fit on degraded images and videos. Another approach employs restoration-based models to improve the quality of compressed images and videos. Specifically, we choose the ``all-in-one'' restoration model PromptIR~\cite{potlapalli2023promptir}, the compression restoration model PromptCIR~\cite{li2023promptcir}, and the recognition-oriented restoration model Cascaded Network (CascadeNet)~\cite{ddnet}. 

\subsubsection{Training Details}

The framework includes codecs, task models, and CDRE-related modules. Video and image codecs remain fixed during training. In ``Ours," CDRE-related and task models are jointly optimized with the default configuration of downstream models; in ``Ours-FD," downstream models are frozen while CDRE-related modules are trained for 60k iterations. (More training details are described in supplementary materials.) Values of $\lambda$ are $4.0$ (video segmentation) and $0.1$ (object detection and keypoint detection). More details are shown in the supplementary materials.


\begin{table}[t]
    \caption{BD-rate(\%)↓ of compared methods on different downstream tasks. Bold indicates the best result. }
    \vspace{-8pt}
    \centering
    \begin{tabular}{|l|c|c|c|}
        \hline
        \multirow{2}{*}{Methods} & Object & Video instance & Keypoints \\
         & detection & segmentation & detection \\
        \hline
        Direct test & 0.0 & 0.0 & 0.0 \\
        \hline
        Head finetune & -8.89 & -32.54 & -6.84 \\
        \hline
        Full finetune & -58.53 & -38.96 & -29.18 \\
        \hline
        PromptIR~\cite{potlapalli2023promptir} & +11.57 & +6.27 & -1.39 \\
        \hline
        PromptCIR~\cite{li2023promptcir} & +3.33 & -3.05 & -3.86  \\
        \hline
        CascadeNet~\cite{ddnet} & -14.39 & -7.00 & -5.18 \\
        \hline
        Ours-FD & -21.70 & -34.95 & -9.83\\
        \hline
        Ours & \textbf{-66.88} & \textbf{-53.68} & \textbf{-30.34}\\
        \hline
    \end{tabular}
    \label{table_bd_rate}
    \vspace{-16pt}
\end{table}

\subsection{Rate-task Perforamnce}

The BD-rate (\%) metric~\cite{dcvc_dc} represents the percentage of bitrate changing while achieving the same task performance compared to the anchor. The anchor is established by directly employing the downstream model on the compressed data. 

\subsubsection{With HVS-oriented Codec}

\begin{figure}[t]
    \centering
    \includegraphics[width=0.95\linewidth]{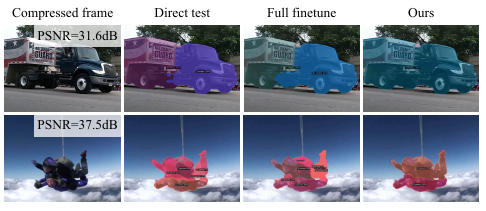}
     \vspace{-10pt}
    \caption{Visualization of video instance segmentation. The above frames are compressed by H.264/AVC with $crf=35$. Our proposed CDRE framework achieves better results after lossy compression. }
    \label{figure_visualization}
    \vspace{-12pt}
\end{figure}

Fig.~\ref{curve_hvs} presents the results of three tasks across various bitrates and codecs. Our method significantly enhances rate-task performance with a minimal increase in bits per pixel (bpp). As shown in Table~\ref{table_bd_rate}, it reduces the BD-rate by 8.35\% (object detection), 14.72\% (instance segmentation), and 1.16\% (keypoint detection) while fine-tuning downstream models. It saves 21.70\% (object detection), 34.95\% (instance segmentation), and 9.83\% (keypoint detection) bitrates when downstream models are fixed. Due to the page limitation, the rate-task curve of keypoint detection is shown in supplementary materials. Visualization on video instance segmentation is shown in Fig.~\ref{figure_visualization}. 

\begin{figure}[!t]
    \centering
    \includegraphics[width=0.90\linewidth]{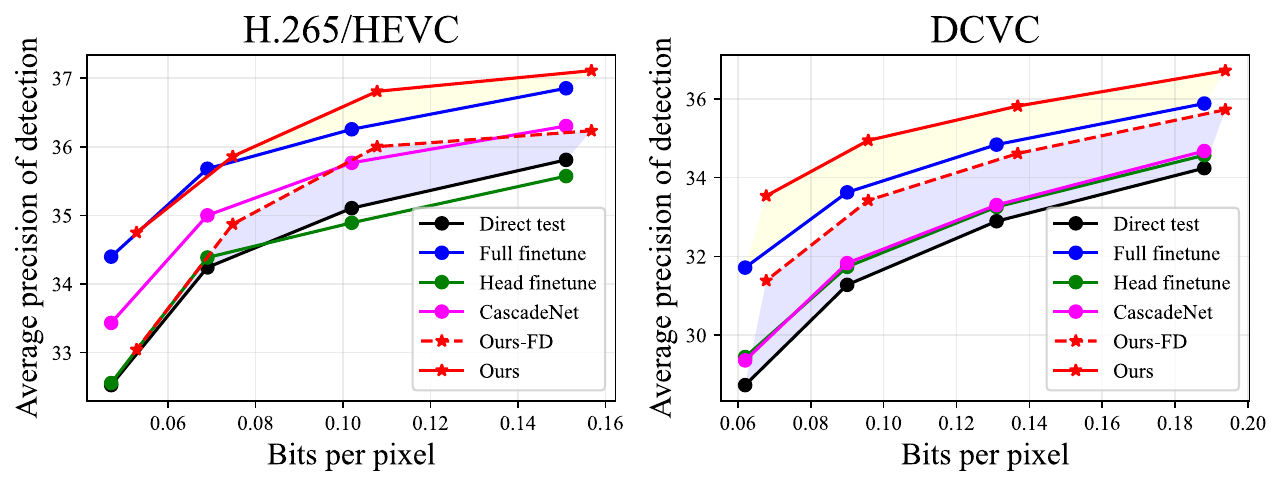}
    \vspace{-12pt}
    \caption{Rate-task performance of the CDRE framework trained on H.264/AVC-compressed data but tested on video compressed by H.265/HEVC and DCVC. }
    \label{curve_transfer}
    \vspace{-8pt}
\end{figure}

To valid the transferability of CDRE, it is trained on H.264/AVC-compressed videos but is tested on videos compressed by H.265/HEVC and DCVC. Fig.~\ref{curve_transfer} shows that CDRE has strong transferability across different codecs and achieves the best performance. 

\begin{figure}
    \centering
    \includegraphics[width=0.90\linewidth]{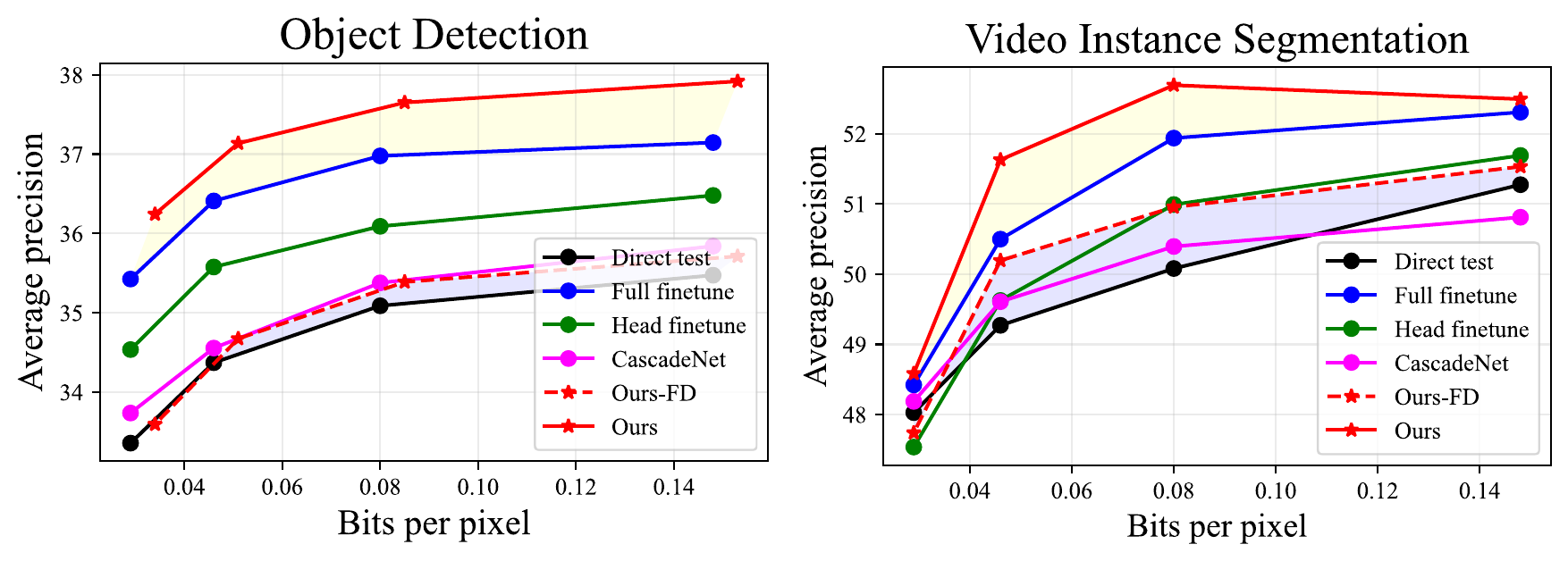}
    \vspace{-12pt}
    \caption{Rate-task performance of CDRE framework with VCM-related codec SMC++ on object detection and video instance segmentation.}
    \label{curve_vcm}
    \vspace{-16pt}
\end{figure}

\subsubsection{With VCM-related Codec}

The CDRE is also tested on VCM-related codec SMC++~\cite{tian2024smc++}. As shown in Fig.~\ref{curve_vcm}, the CDRE further reduces the bitrate by 26.70\% (object section) and 12.89\% (instance segmentation) with fine-tuning downstream models. It saves 10.48\% and 34.95\% bitrates when downstream models are frozen. 

The above experiments demonstrate that the CDRE effectively improves the rate-task performance of existing codecs. In contrast, training the model on compressed data and restoration-based methods struggles to achieve satisfactory overall performance. Notably, image restoration for HVS-metrics (PromptIR and PromptCIR) offers few benefits to downstream task performance.

\subsection{Overhead Analysis}

The proposed CDRE framework is lightweight. The CNN-based version of the CDRE contains 1.9M parameters, while the Transformer-based version introduces 3.7M parameters. Notably, both versions share the same structure on the encoder side. The encoder part consists of only 18K parameters. For a 720P frame, the CDRE incurs only an additional 6 ms of computation time on NVIDIA RTX 3090. For 720P 30fps video, the additional distortion information bitstream requires about 20KBps, allowing for more bitrate savings of video streams while maintaining machine perception performance. 

The MACs per pixel for each component are 0.89K on the encoder side, 0.33K for the Transformer-version decoder, and 1.61K for the CNN-version decoder. The MACs per pixel of CDRE is less than $2\perthousand$ of the video compression process (e.g., DCVC-DC~\cite{dcvc_dc} has 1.27M, DCVC~\cite{li2021dcvc} has 1.09M).

\begin{table}[!t]
    \caption{Ablation study on compression-sensitive extractor. The anchor of bitrate is Ours.}
    \vspace{-8pt}
    \centering
    \setlength{\tabcolsep}{5pt}
    \begin{tabular}{|c|c|c|c|}
        \hline
        Multi-scale & Cosine Similarity & Modulation & Bitrate increase (\%) \\
        \hline
        \Checkmark & \XSolidBrush & \Checkmark & +5.77 \\
        \hline
        \XSolidBrush & \Checkmark & \Checkmark & +5.13 \\
        \hline
        \XSolidBrush & \XSolidBrush & \Checkmark & +8.14 \\
        \hline
        \Checkmark & \Checkmark & \XSolidBrush & +3.71 \\
        \hline
    \end{tabular}
    \label{table_ablation_extractor}
    \vspace{-12pt}
\end{table}

\begin{table}[!t]
    \caption{Ablation study on progressive distortion representation transformation and embedding. Bold indicates the best result. The anchor of BD-rate is direct test.}
    \vspace{-8pt}
    \centering
    \setlength{\tabcolsep}{9.5pt}
    \begin{tabular}{|c|c|c|c|c|}
        \hline
        Embedding layers & 1 & 2 & 3 & 4 \\
        \hline
        BD-rate (\%)↓ & -62.66 & -62.86 & -63.11 & \textbf{-66.88} \\
        \hline
    \end{tabular}
    \label{table_embedding}
    \vspace{-12pt}
\end{table}

\begin{table}[!t]
    \caption{Ablation study on channels of compact representation. Bold indicates the best result. The anchor of BD-rate is direct test}
    \vspace{-8pt}
    \centering
    \setlength{\tabcolsep}{7pt}
    \begin{tabular}{|c|c|c|c|c|c|}
        \hline
        Channel count & 1 & 3 & 6 & 10 &16 \\
        \hline
        BD-rate (\%)↓ & -64.14 & -62.53 & \textbf{-66.88} & -59.66 & -55.49 \\
        \hline
    \end{tabular}
    \label{table_dim}
    \vspace{-12pt}
\end{table}

\subsection{Ablation Study}

\subsubsection{Compression-sensitive Extractor} 

As shown in Fig.~\ref{table_ablation_extractor}, to verify the effectiveness of amplifying distortion in the feature domain, cosine similarity is removed, leading to a 5.77\% bitrate increase. Moreover, to assess the impact of multi-scale distortion analysis, we simplified the compression-sensitive extractor to a single scale, resulting in a 5.13\% bitrate increase. When both ablations were applied, the bitrate increased by 8.14\%. Additionally, linear modulation enables the distortion encoder to better leverage feature distortion, reducing the bitrate by 3.71\% compared to simple concatenation. All ablation studies are conducted on object detection. 

\subsubsection{Distortion Representation Embedding}

To explore the effect of progressive distortion representation transformation and embedding, we change the number of embedded layers. As shown in Table~\ref{table_embedding}, introducing more layers of transformation and embedding enables the downstream models to better leverage distortion information. 

\subsubsection{Influence on the Dimension of Compact Representation}

Although transmitting and embedding distortion representation can improve task accuracy, it also introduces bitrate overhead. The dimension of compact representation must strike a balance between accuracy gain and bitrate cost. To address this, different numbers of channels are tested. As shown in table~\ref{table_dim}, the best rate-task performance is achieved with 6 channels. 

\section{Conclusion}

In this paper, we propose a Compression Distortion Representation Embedding (CDRE) framework to enhance the rate-task performance of existing codecs. We develop a compression-sensitive extractor to amplify and identify distortion in the feature domain. Moreover, a lightweight distortion codec is introduced to compress distortion information into a compact representation for effective transmission. Additionally, we introduce a distortion representation embedding module that progressively transforms and embeds the distortion feature into downstream models, making them aware of compression degradation of the current input. Our framework is evaluated on three downstream tasks—object detection, video instance segmentation, and person keypoint detection—across seven codecs. Experimental results show that the proposed CDRE framework significantly improves the rate-task performance with minimal overhead in terms of bitrate, execution time, and number of parameters. 

\bibliographystyle{IEEEbib}


\end{document}